\documentclass[fleqn,10pt]{wlscirep}
\title{Emergence of leader-follower hierarchy among players in an on-line experiment}

\author[1,]{B{\'a}lint J.\ T{\'o}th}
\author[2,*]{Gergely Palla}
\author[3]{Enys Mones}
\author[4]{Gerg{\H o} Havadi}
\author[1]{N{\'o}ra P{\'a}ll}
\author[2]{P{\'e}ter Pollner}
\author[1]{Tam{\'a}s Vicsek}
\affil[1]{Dept.\ of Biological Physics, E{\"o}tv{\"os} University, H-1117 Budapest, P{\'a}m{\'a}ny P.\ stny.\ 1/A, Hungary}
\affil[2]{MTA-ELTE Statistical and Biological Physics Research Group, H-1117 Budapest, P{\'a}m{\'a}ny P.\ stny.\ 1/A, Hungary}
\affil[3]{Dept.\ of Applied Mathematics and Computer Science, Technical University of Denmark, Richard Petersens Plads 324, DK-2800 Kgs. Lyngby, Denmark}
\affil[4]{Hungarian Academy of Sciences, Centre for Social Sciences "Lend{\"u}let" Research Center for Educational and Network Studies (RECENS), T{\'o}th K{\'a}lm{\'a}n u. 4. H-1097 Budapest, Hungary}
\affil[*]{pallag@hal.elte.hu}


\begin{abstract}
Hierarchical networks are prevalent in nature and society, corresponding to groups of actors - animals, humans or even robots - organised according to a pyramidal structure with decision makers at the top and followers at the bottom. While this phenomenon is seemingly universal, the underlying governing principles are poorly understood. Here we study the emergence of hierarchies in groups of people playing a simple dot guessing game in controlled experiments, lasting for about 40 rounds, conducted over the Internet. During the games, the players had the possibility to look at the answer of a limited number of other players of their choice. This act of asking for advice defines a directed connection between the involved players, and according to our analysis, the initial random configuration of the emerging networks became more structured overt time, showing signs of hierarchy towards the end of the game. In addition, the achieved score of the players appeared to be correlated with their position in the hierarchy. These results indicate that under certain conditions imitation and limited knowledge about the performance of other actors is sufficient for the emergence of hierarchy in a social group.
\end{abstract}

\begin{document}

\flushbottom
\maketitle

\thispagestyle{empty}


\section*{Introduction}
In nature, humans and many kinds of animals prefer to form networks with a hierarchical structure. This is supported by several empirical studies, focusing on the dominant-subordinate hierarchy among crayfish \cite{Huber_crayfish}, the leader-follower network of pigeon flocks \cite{Tamas_pigeons}, the rhesus macaque kingdoms \cite{McCowan_macaque}, and social interactions \cite{Guimera_hier_soc,our_pref_coms,Sole_hier_soc}. In addition to systems composed of interacting living species, signs of hierarchy were observed also  in the transcriptional regulatory network of Escherichia coli \cite{Zeng_Ecoli}, in neural networks \cite{Kaiser_neural} and  technological networks \cite{Pumain_book}, among scientific journals \cite{Palgrave}, in urban planning \cite{Krugman_urban,Batty_urban}, ecological systems \cite{Hirata_eco,Wickens_eco}, and evolution \cite{Eldrege_book,McShea_organism}. A serious advantage of hierarchical organisation in case of a group of individuals is that it maximises the group performance under certain conditions \cite{Anna_nature,Tamas_Maryam_hier,Anna_Tamas_hierbook}. 

Another plausible advantage of hierarchy is that it enables an effective information flow in the network and thus, may improve collective decision making. In the last decade, considerable attention has been paid on the description and characterisation of collective intelligence and the wisdom of the crowd \cite{Surowieczki}. The widespread applications of collective intelligence already contribute to the 21-century global economy, for example crowd-sourcing is applied to tasks ranging from translation between natural languages through improvement of optical character recognition systems to forecasting future events.

A relevant question regarding hierarchical agent groups is related to the emergence of hierarchy itself, i.e., how can an initially random configuration of the inter relations become hierarchical, and what are the necessary conditions for the emergence of a hierarchy? Quantitative studies of interacting agents often relies on the toolbox provided by game theory. Game theory itself is centred around the notions of costs and benefits (individual or mutual) of cooperative or antagonistic relations, but does not describe the flow of information \cite{Holme_PRL,Newman_hier,Turchin_hier_soc}. We argue that hierarchy can be the result of imitation, as a very common, yet less studied manifestation of social interactions. Copying the behaviour of better performing individuals is a natural strategy, which can be effective for achieving one's goals within communities of varying sizes ranging from e.g., a high school class of a sport club through larger organisations and institutions to the society as a whole itself. In addition to humans, imitation is known to be present to some extent in groups of enculturated apes as well \cite{chimpanze,enculturated_apes,human_and_ape_imitation}, and according to experiments on homing pigeons, copying the flight direction of better navigating flock members seemed to be a widespread strategy amongst the pigeons, which leaded to the emergence of a leader-follower hierarchy in the flock \cite{Tamas_pigeons}. 

Model simulations of agents (which could represent people, animals, drones, etc.), who needed to carry out a simple task and were allowed to copy from each other, showed that under certain conditions the network between agents became hierarchical over time \cite{Nepusz_2013}. Here we show that imitation between individuals can lead to the emergence of a hierarchy in a real network between human players as well.

\begin{figure}[h]
\centering
\includegraphics[width=0.55\textwidth]{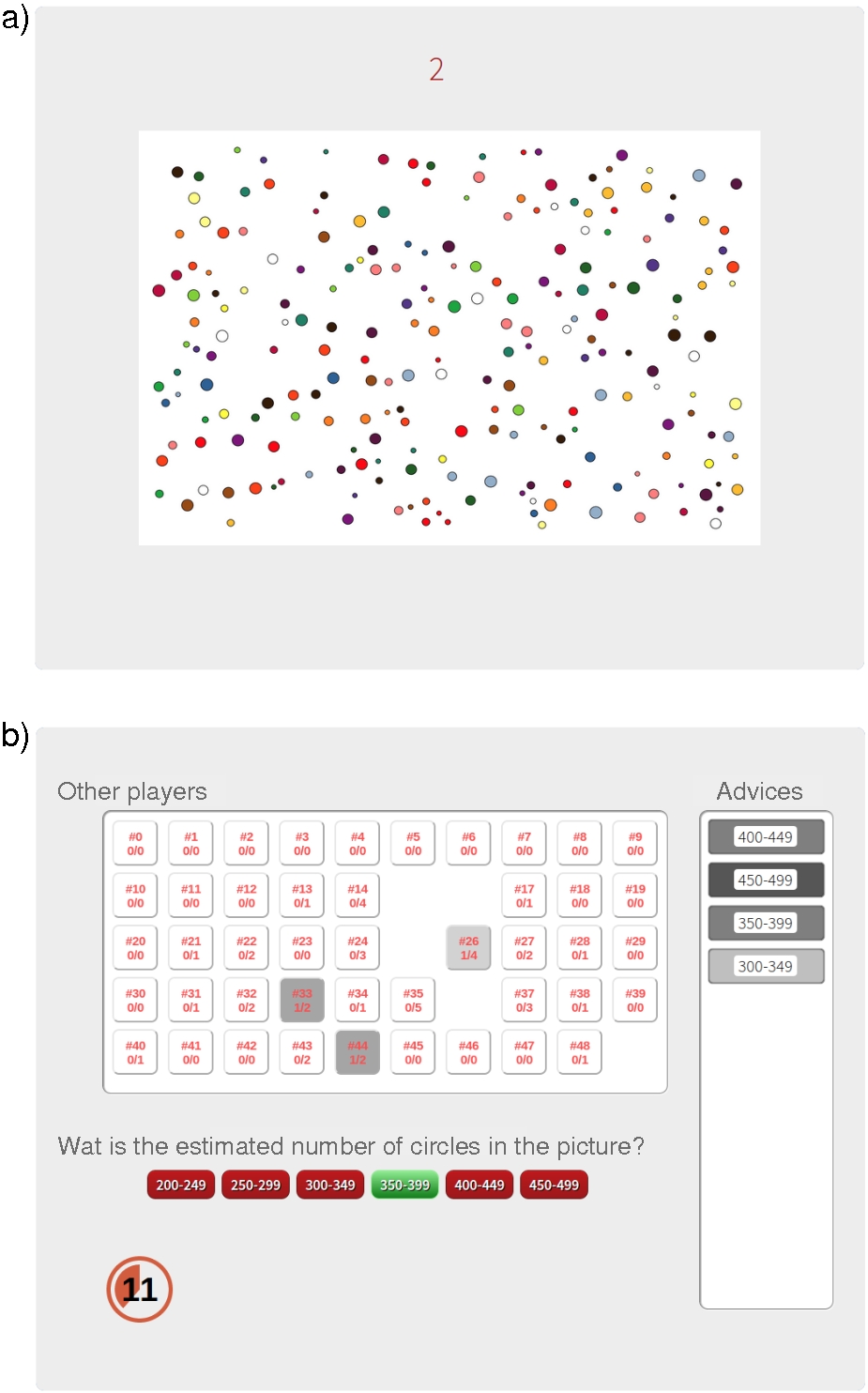}
\caption{ Screen-shots from the dot-guessing game. a) An actual exercise. b) Inquiry of advises. }
\label{fig:schot}
\end{figure}

We have designed a gamefied experiment (under the codename Picturask) that we conducted over the Internet. The subjects participated in a game, where everyone had to approximate the number of circles on an image in subsequent rounds, like shown in Fig.\ref{fig:schot}. Before making the next guess, players were able to see the others previous guesses through clicking on other players' tiles. Since the number of circles in the image was changing with a relative low frequency, these actions can be treated as asking for advice from the other players, and they define a time evolving network of relations between the players. The player's action through the game has been recorded, and these records were later used to inference the behaviour of players. The results indicate that the interaction networks between the users were becoming hierarchical towards the end of the games, roughly after 30-40 rounds. After the game, also deep interviews has been conducted with the players to see if our observations are in agreement with the players' intentions. 

\section*{Methods and data}
\subsection*{Subjects}
Altogether 170 users signed up for the game of which 96 participated in more than $90\%$ of the turns. The majority ($89\%$) of the players were of age 18 to 28, thus $65\%$ of them received at least bachelor's degree or equivalent, and 33\% accomplished high school. $56\%$ of the players were female. Subjects were divided into three groups to play the game, each group playing on different weeks, for three consecutive days. Participation was encouraged by a 2000 HUF reward. (6.7 EUR, approximately $1\%$ of average salary in Hungary) Four players with the best results was granted special awards as motivation. 

\subsection*{Procedure}
Subject participated in our experiment online from home, or whereas they had Internet access. The game was based on a standard LAMP architecture, and players were able to participate through their favourite web browsers. During the game the participants were asked to accomplish a simple estimation: they had to guess the number of bubbles on randomly generated images. As the game advanced, the correct answer changed with a relatively slow frequency. (Once in five turn, even though this was unknown by the players.) The total length of the game was 40 rounds, lasting for about an hour. Whilst answering, the player could reveal the previous answer of a maximum of 10 other players in each turn by clicking on its tile. This act is called asking for advice, and the player being asked is called adviser. The motivation for this process was to reproduce the agent-based simulations showed in \cite{Nepusz_2013}.

Depending on the ratio of advises that turned out to be correct, the tiles of the rivals got darker, highlighting the advisers who gave the player the best advises. The estimated knowledge (the ratio of correct advises) was also displayed on the rival tile, but no other information was available. In fact, the rivals were displayed in a random permutation for each player in order to avoid potential biases introduced by the order of the tiles. (However, the permutation of the other players was kept fixed on the display during a game). 

We ran altogether nine experiments, the number of participants in the individual games varied between 30 and 50. Two games served as control experiments, where the act of revealing the previous answer of another player returned a random sample from the previous answer of all other players. Naturally, users were not aware of the different behaviour of their interface during the control experiments, and assumed normal functioning. Nevertheless, under these circumstances the observed quality of the other players advice tends to become homogeneous as the game is advancing, and the ``best'' adviser on the player's interface is corresponding to a random choice between the other players.

\subsection*{Data}
The datasets supporting this article have been uploaded as part of the Supplementary Information. 

\section*{Results}
\subsection*{Emergence of hierarchy}
During the analysis of the game records we represented the social structure between the players as a time evolving directed network, where links are pointing from the source of the advice to the person who has asked for the advise. Naturally, the weight of a connection at a given round $r$ is given by the number of times the target user of the link has asked for advice from the source of the link. Our primary conjecture about the experiments was that since the less-well performing players would prefer to ask advice from  better performing individuals, the links towards the best advisers would be reinforced during the game-play, and the participants would voluntarily arrange themselves into a hierarchical network. Since the information in this network is flowing from the player who was asked to the player who was asking, we considered the links to be pointing always in the direction of the players asking for advise.

Quantifying the importance of hierarchical organisation in the structure of a network is non-trivial problem with a number of possible different approaches available \cite{Enys_2012,Luo_Magee,Sole_hier_PNAS,RWH}. A natural idea is to introduce a hierarchy measure, which can be viewed as a function on the domain of graphs,  $H:\mathbb{G}\mapsto\mathbb{R}$, mapping a graph $\mathcal{G}\in\mathbb{G}$ into a real number, $H(\mathcal{G})\in\mathbb{R}$. The value of the measure is actually $H(\mathcal{G})\in[0,1]$ or $H(\mathcal{G})\in[-1,1]$ in most cases, with high values corresponding to hierarchical structures and low values indicating the absence of hierarchy in the examined network.

Here we adopt the hierarchy measure named as the Global Reaching Centrality, $H_{\rm GRC}$, which is based on the in-homogeneity of the reach distribution of the nodes \cite{Enys_2012}, and already proved to be an intuitive and successful approach in a number of studies \cite{Shimoji20140599,Enys_Peter_Tamas,Palgrave}. The $m$-reach of a node $i$,  denoted by $C_m(i)$ is given by the fraction of nodes that can be reached from $i$ in at most $m$-steps. In a hierarchical network $C_m(i)$ is expected to be high for nodes at the top of the hierarchy, whereas $C_m(i)$ is assumed to be low for the bottom nodes. In contrast, for a non-hierarchical network (such as a random graph), $C_m(i)$ is more or less the same for all nodes. Based on that, $H_{\rm GRC}$ is defined by quantifying the in-homogeneity of the $m$-reach distribution as
\begin{equation}
H_{\rm GRC}=\frac{\sum_{i=1}^N\left[ C_m^{\rm max}-C_m(i)\right]}{N-1},
\label{eq:H_GRC}
\end{equation}
where $C_m^{\rm max}$ is the maximum of the $m$-reach. 

Although the original definition of $H_{\rm GRC}$ in Ref.\cite{Enys_2012} was parameter-free, corresponding to $m=\infty$, here we stick to the above version where $m$ is a tunable parameter. The reason for this is that the networks between the players are relatively small, and from a significant part of the nodes we can reach the entire network if there is no restriction on the length of the paths we can take. Therefore, if $m$ is larger than the diameter of the network, $C_m(i)$ becomes maximal for these nodes, and we can no longer distinguish between them based on $C_m(i)$ (which also means that we cannot tell who is higher in the hierarchy, and who is lower). According to that, $m$ should be chosen low enough to prevent the saturation of $C_m(i)$ for the majority of the nodes, and allowing a maximal reach for only the very top nodes in the hierarchy. 

In addition to $H_{\rm GRC}$ we also calculate the value of a much simpler hierarchy measure given by the Link Flow Hierarchy, $H_{\rm LFH}$ in each round. This hierarchy measure is corresponding to the fraction of links not taking part in any directed cycle \cite{Luo_Magee},
\begin{equation}
H_{\rm LFH}=\frac{M-M_{\rm cyc}}{M},
\label{eq:H_LFH}
\end{equation}
where $M_{\rm cyc}$ denotes the number of links that are included in at least one directed cycle, and $M$ is corresponding to the total number of links in the network.

Before actually applying (\ref{eq:H_GRC}) and (\ref{eq:H_LFH}) to the time evolving network between the players, we need to take into consideration the following two remarks. First, these hierarchy measures do not take into account the link weights, therefore, the information in e.g., the reinforcement of relevant connections is lost. Second, both $H_{\rm GRC}$ and $H_{\rm LFH}$ (together with basically all other hierarchy measures in the literature) are sensitive to the overall link density of the network. In our case the number of links can only increase during the games, thus, the overall link density is also increasing. 

In order to treat these problems, in each round we sorted the in-links on every node according to their current weight (corresponding to the number of times advice was asked from the corresponding neighbour up till the given round), and kept only the first $k$ during the calculation of the hierarchy measure. On the one hand this ensures that the overall link density becomes roughly constant during the game, except for a short transient at the beginning of the game, under which the first links become established. On the other hand, this way we also take into account the link weights in an implicit form, as stronger connections have a higher likelihood to be in the top $k$ links of their target compared to weak links. According to that, if a connection is reinforced by repeated ``asking for advice'' actions between the same players over multiple rounds, then consequently its chance to be included in the evaluation of (\ref{eq:H_GRC}) and (\ref{eq:H_LFH}) becomes higher compared to a connection over which only a single or a few advises were asked. 

Whenever a draw between the weights of competing in-links on the same player occur, (which happens rather often), the choosing of the first $k$ in-neighbours must involve a random selection from multiple links of same strength. This enables us to define an ensemble of evolving networks, where each member is corresponding to a possible realisation of the random choices mentioned above.  The advantage of this approach is that the network characteristics of interest, (such as hierarchy measures), can be evaluated over the ensemble instead of only on a single empirically observed network, and therefore, beside the value of the given characteristic we can also estimate its uncertainty within the ensemble.

\begin{figure}[h]
\centerline{\includegraphics[width=0.8\textwidth]{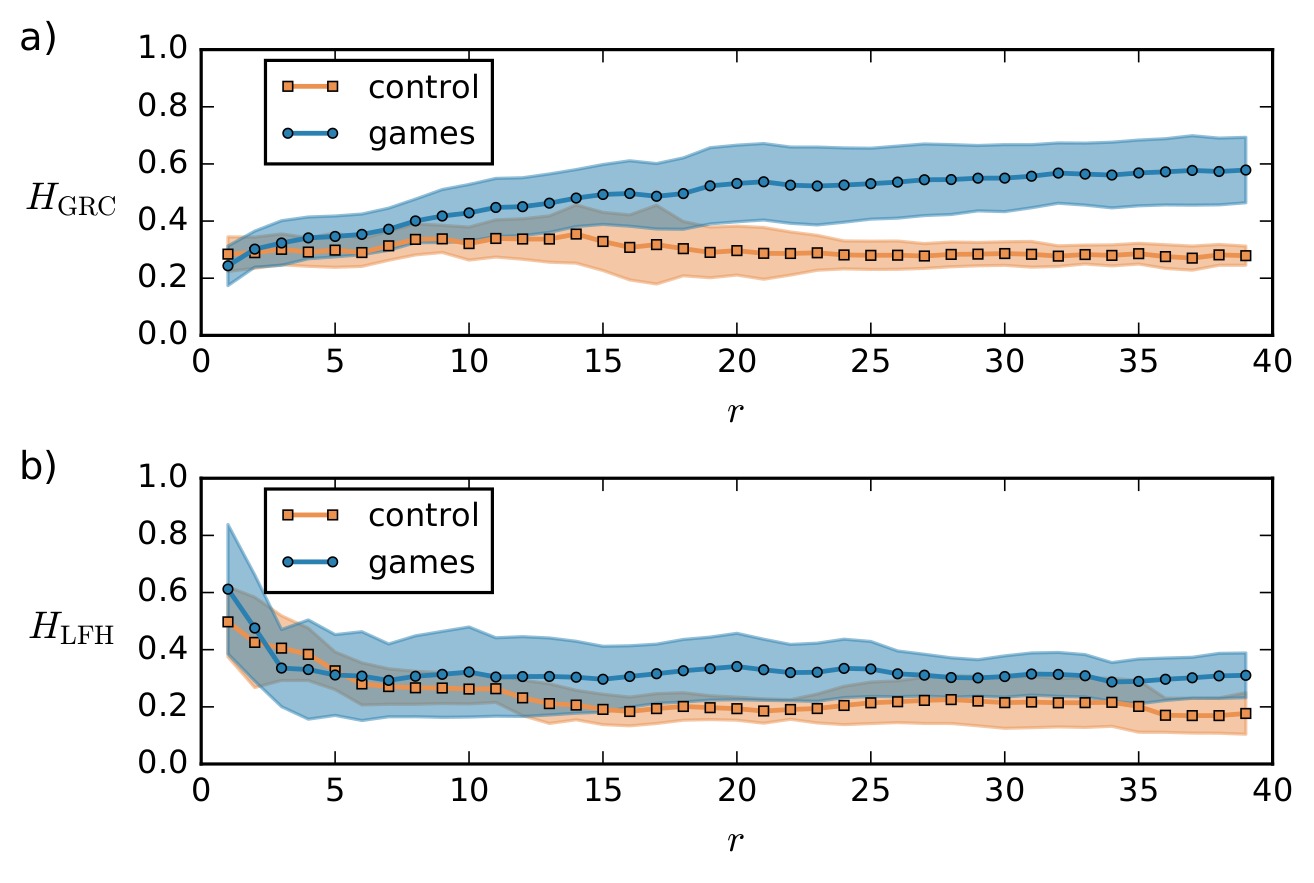}}
\caption{ The evolution of hierarchy measures during the experiments. a) The global reaching centrality, $H_{\rm GRC}$, as a function of the number of rounds $r$ during the games (blue) and during the control experiments (orange). Points connected by a solid line are corresponding to the average, whereas the shaded area indicates the standard deviation around the average. b) The link flow hierarchy $H_{\rm LFH}$ as a function of the number of rounds, plotted in the same manner as $H_{\rm GRC}$ in panel a).}
\label{fig:hier_evolv}
\end{figure}

In Fig.\ref{fig:hier_evolv}.\ we show the time evolution of the hierarchy measures during the games and the control experiments. The $k$ parameter corresponding to the maximally allowed in-degree for the nodes was set to $k=3$ and the $m$ parameter in the calculation of the $m$-reach of the players was set to $m=2$. (According to studies detailed in the Supplementary Material, our results are robust against changes in these parameters). For every game in each round we evaluated the hierarchy measures for 500 samples from the network ensemble, in Fig.\ref{fig:hier_evolv}.\ we plotted the average and standard deviation of these for the real games in blue and for the control experiments in orange.
In case of the Global Reaching Centrality (shown in Fig.\ref{fig:hier_evolv}a), we can observe a clear separation between the two curves. The average value of $H_{\rm GRC}$ is showing an increasing tendency in case of the normal games, whereas it stays more or less constant (with some fluctuations) during the control experiments. At the end of the game the gap between the two curves becomes quite significant. Although the average hierarchy is higher for the normal games also in case of the Link Flow Hierarchy (shown in Fig.\ref{fig:hier_evolv}b), the two curves stay more close to each other. The decreasing tendency of the results in the first few rounds is most likely caused by the increase of the average link density in this transient regime. 

\begin{figure}[h]
\centerline{\includegraphics[width=0.8\textwidth]{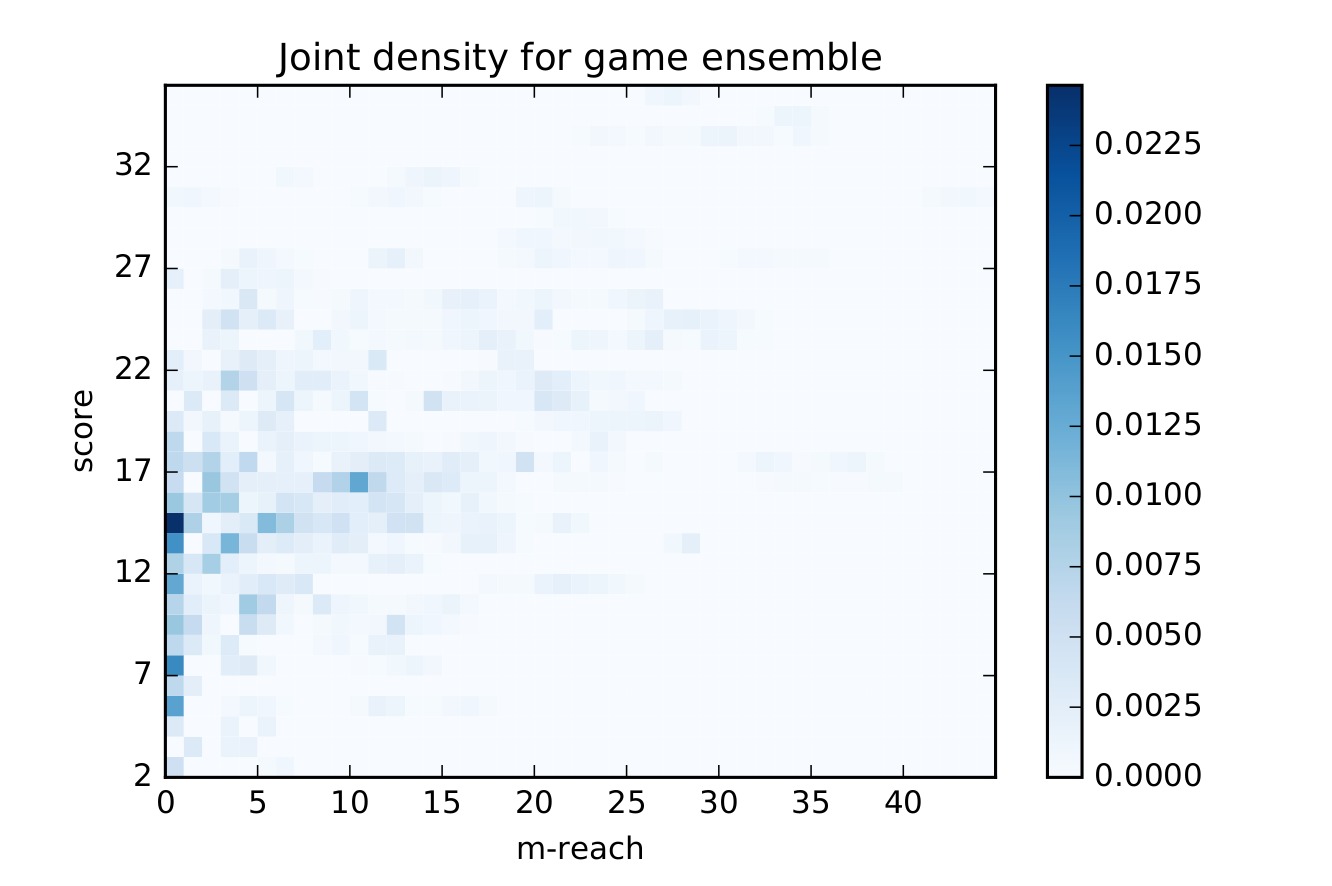}}
\caption{ The joint probability density of the $m$-reach and the score of the players for the normal games, where both the $m$-reach and the score was evaluated at the end of the game. The probability of a given cell (corresponding to a given pair of $m$-reach and achieved score values) is indicated by the colour heat map.}
\label{fig:games_heat_map}
\end{figure}
\begin{figure}[h]
\centerline{\includegraphics[width=0.8\textwidth]{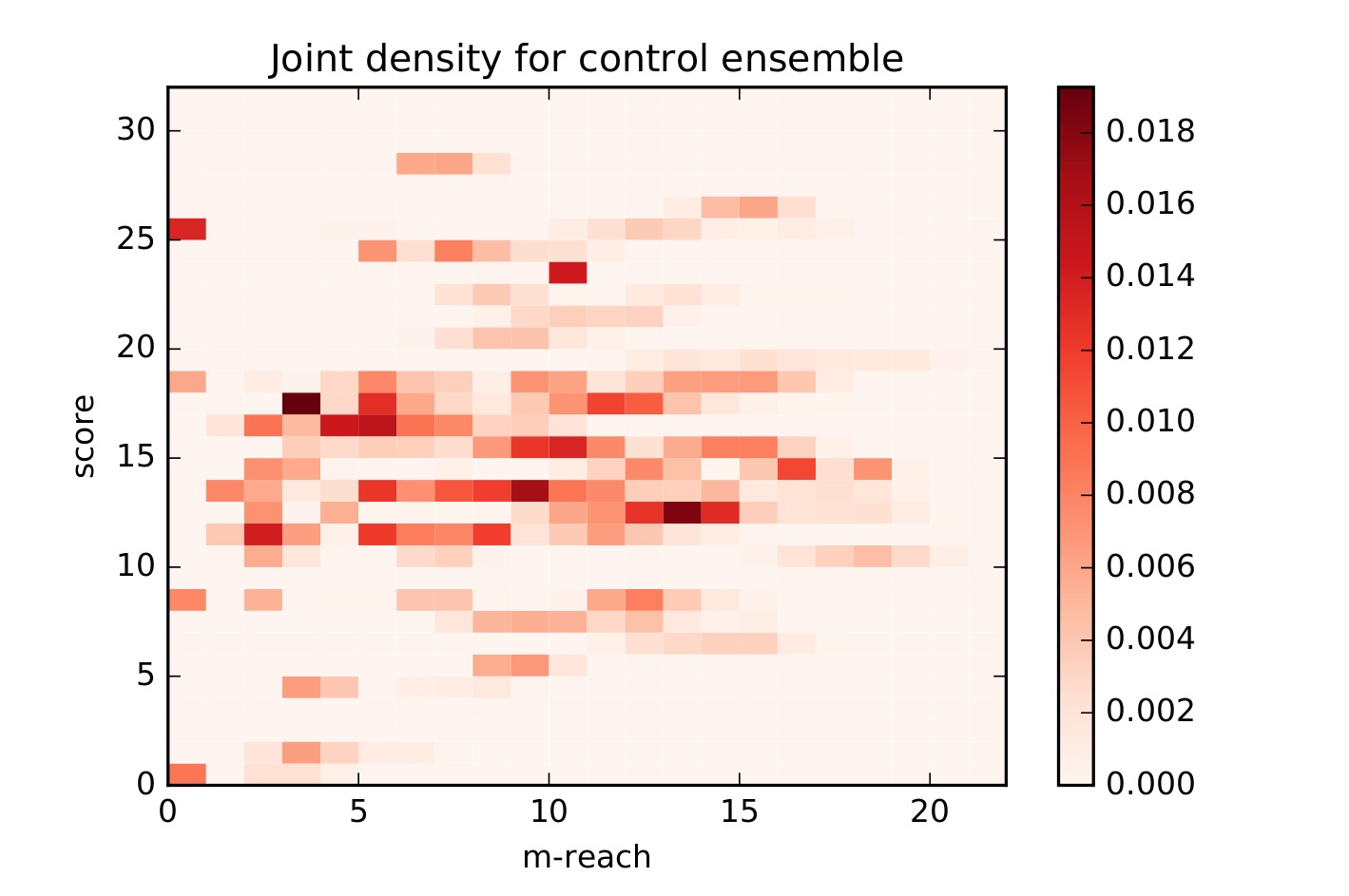}}
\caption{ The joint probability density of the $m$-reach and the score of the players for the control experiments, plotted in the same way as in Fig.\ref{fig:games_heat_map} for the real games.}
\label{fig:control_heat_map}
\end{figure}
In Figs.\ref{fig:games_heat_map}-\ref{fig:control_heat_map}.\ we show the relation between the position of the players in the hierarchy and their achieved scores by plotting the joint probability distribution between the $m$-reach and the scores. For the normal games (Fig.\ref{fig:games_heat_map}.) the heat map suggests that players higher in the hierarchy (having a larger $m$-reach) will more likely achieve a higher score compared to individuals at the bottom of the hierarchy. In contrast, these two quantities seem to be more or less independent in case of the control experiments (Fig.\ref{fig:control_heat_map}.). These observations are supported by the Pearson correlation coefficients which turned out to be $C=0.518$ in case of the normal games and only $C=0.138$ for the control experiments.

\begin{figure}[h]
\centerline{\includegraphics[width=0.8\textwidth]{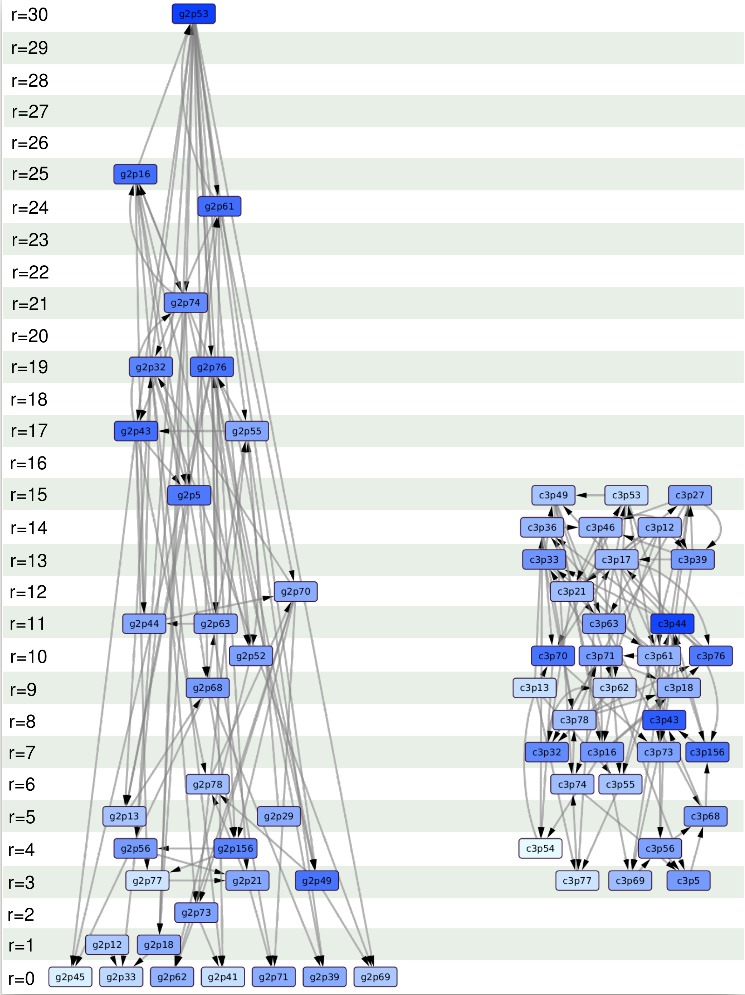}}
\caption{Samples from the network ensembles, one from the normal games (left) and one from the control experiments (right). The vertical position of the nodes is determined by their $m$-reach, whereas their colour is indicating the achieved score at the end of the game (where darker shades are corresponding to higher scores).}
\label{fig:net_illustr}
\end{figure}

Finally, in Fig.\ref{fig:net_illustr}.\ we also show two samples from the network ensembles for illustration. The graph on the left was chosen from the networks at the final round of a normal game, whereas the network on the left was sampled from the network ensemble obtained at the final round of a control experiment.
The position of the nodes in the layouts reflects their $m$-reach, whereas the colouring indicates the achieved score at the end of the game. The difference between the two networks is quite apparent. The range of the $m$-reach is more than twice as large in case of the normal game compared to the control experiment, spanning from $0$ to $30$ in the first case, while being limited only between $3$ and $15$ in the second case. Furthermore, the achieved scores and the position in the hierarchy according to the $m$-reach are apparently correlated in case of the normal game (where top nodes tend to obtain better scores), whereas this is not the case in the control experiment, where the achieved score seems to be independent of the node position.

These results suggest that the network between the players is evolving in different manner compared to the control experiments, and the self-organisation of the connections is leading to the a more hierarchical network structure towards the end of the game compared to the random graphs observed during the first few rounds. It is worth mentioning that although the emerging networks are clearly different in the real games and in the control experiences, the vast majority of players did not notice any difference. The rate as the set of advisers stabilised were practically the same in all the nine games. We believe this is an indication of the existence of a characteristic time that is necessary to discover differences in performance, otherwise the social structure remains random.

\subsection*{The decision making process}
Based on the interviews, there are three main stages of decision making in Picturask: first the player gets an initial guess, then he collects information from others, and finally she makes her final decision using the information collected from others. It is important to separate information gathering from the decision making process.

The only information a player knows about the others is the colour of the pile that reflects the ratio of correct advises. Even if the player doesn't understand what the exact meaning of the pile is, she can use it as a heuristic: she links good performance with the colour of the pile. So the colour has two functions: in most cases, it indicates the pool of possible advice givers and the order of asking advice (the darker is better) and rarely, it can also give more weight to the advice of some players. Players use the collected information in three ways: they verify their initial guess, it helps them choose between different alternatives or they simply take the majority's guess as their answer.

\section*{Discussion}
From social psychological aspects the players in the game form a small group which can be characterised by interdependence where individuals act in a common interpersonal space, while they influence each other{'}s actions in a special way \cite{Johnson2}. In our research, this space is somewhat artificial because it is determined by the rules of the game; for example direct communication is not allowed: players can only see the others{'} anonymised guesses. It depends on the player whether she considers and/or accepts the tip as an advice.

This process is similar to social exchange theory \cite{Cropanzano}, where the subject of this interaction is the information about each other{'}s tips. Players gain this information from the colour of the tiles, where darker colours signal better results in the former rounds. Therefore, tiles can be identified as heuristics. The theory of heuristics describes the stereotyping pattern of human being where a complex question may be often answered with a simplifying method {\textendash} like players tend to judge a tip based on the colour of tiles \cite{Gilovich_2002}.

When designing our experiment one of our primary concerns was to remain consistent with the model presented in \cite{Nepusz_2013}. This resulted in less intuitive set of rules, thus - according to the interviews - not every player was clear about the fine details of the game. 
However we believe these circumstances do not affect our final conclusions. 

In summary we studied the emergence of hierarchy in the social network between competing players of a simple online dot guessing game in controlled experiments. According to our results, the hierarchy measures in the observed networks showed an increasing tendency on average as a function of the number of rounds, when the players could reinforce their connections to other players who have given useful advise in the previous rounds. It is important to note that the players were not aware of the overall score of the other players, the only information they received was given by the fraction of correct advises from the previously asked other players, (indicated by the colour of their tile). During the control experiments, where the advise from the other players was randomised, the hierarchy measures remained constant or showed a decreasing tendency as a function of the number of rounds. This is consistent with the fact that under this setting, the connections between the players are expected to be random. 

These results indicate that  imitation and limited knowledge about the performance of other actors is sufficient for the emergence of a hierarchy. Imitation is a behaviour that is widespread in the human society and is known to be present in groups of various other species as well. Similarly, hierarchical organisation is also a prominent characteristic of networks representing the inter relations between group members in case of numerous different species (including humans as well). Our research has shown that these two relevant cross-species phenomena can be easily in causal relation with each other.


\section*{Acknowledgements}

The authors are grateful to {\'A}gnes Buv{\'a}r for the fruitful discussions on the sociological aspects of the research. The project has received partial funding from the European Commission's Seventh Framework Programme for Research under EU FP7 ERC COLLMOT, Grant No: 227878 and from the European Union's Horizon 2020 research and innovation programme under grant agreement No. 740688 (RedAlert).

\section*{Author contributions statement}

T.V. designed the experiment, B.T. programmed the LAMP architecture, E.M., G.H., N.P. and B.T. ran the experiments and collected the data, G.P., P.P., E.M. and B.T. analysed the data and performed the statistical analysis, G.P., T.V., P.P. and B.T. wrote the paper with substantial input from all co-authors.





\end{document}